\documentclass[twocolumn,]{aastex631}

\usepackage{graphicx}	
\usepackage{amsmath}	
\usepackage{hyperref}
\usepackage{cprotect}

\usepackage{needspace}
\usepackage{etoolbox}
\pretocmd{\section}{\Needspace{4\baselineskip}}{}{}
\pretocmd{\subsection}{\Needspace{3\baselineskip}}{}{}
\pretocmd{\subsubsection}{\Needspace{3\baselineskip}}{}{}

\newcommand{\jr}{\color{black}}

\begin{document}

\title{Testing the prevalence of hydrogen-silicate miscibility in young sub-Neptunes}

\author[0000-0001-7615-6798]{James G. Rogers}
\affiliation{Institute of Astronomy, University of Cambridge, Madingley Road, Cambridge CB3 0HA, United Kingdom}

\author[0000-0002-0298-8089]{Hilke E. Schlichting}
\affiliation{Department of Earth, Planetary, and Space Sciences, The University of California, Los Angeles, 595 Charles E. Young Drive East, Los Angeles, CA 90095, USA}



\begin{abstract}
Hydrogen-silicate miscibility can significantly alter the interior structure and thermal evolution of sub-Neptunes. We consider the interplay between this miscibility and stellar-driven atmospheric escape. We find that, for the first $\sim 100$~Myrs, sub-Neptunes store most of their hydrogen content within their miscible interiors, protecting it from escape. As hydrogen is removed from the top of the atmosphere, more hydrogen is exsolved from the miscible interior, resupplying the envelope mass and delaying the planet's contraction when compared with models that do not account for miscibility. We construct a population-level Bayesian hierarchical inference model which tests for the prevalence of fully miscible sub-Neptunes. With conservative measurement uncertainty and no completeness corrections, we predict that $\sim 100-200$ observed young sub-Neptunes with measured radii, masses and ages $\lesssim 1$~Gyr are required to differentiate between fully miscible and immiscible models. This survey size requirement decreases with improved measurement uncertainty. Furthermore, one can reliably infer the underlying distributions of planetary initial conditions if completeness corrections are included in the analysis.
\end{abstract}



\section{Introduction} \label{sec:intro}
Observations of young exoplanets are extremely valuable for our understanding of planet formation. Planets can retain the imprints of their formation history during the first $\sim10- 100$~Myr of life, before being sculpted by complex evolutionary processes that mask these signatures over the following $\sim$~Gyr of evolution where typical observations occur.

The value in amassing young planet observations is particularly important for the population of low mass ($\lesssim 20 M_\oplus$), close-in exoplanets (periods $\lesssim 100$~days) referred to as super-Earths and sub-Neptunes. Such planets are thought to be the most common class of planet in our galaxy \citep[e.g.][]{BoruckiKeplerII,Fressin2013,Petigura2013}. After $\sim 1-10$~Gyr of evolution, super-Earths and sub-Neptunes are observed to form a bimodal size distribution, often referred to as the radius valley, with peaks at $\sim 1.5 R_\oplus$ and $\sim 2.4 R_\oplus$, respectively \citep[e.g.][]{Fulton2017,VanEylen2018,Berger2020,Petigura2022,Ho2024}. It is widely believed that the radius valley is a product of atmospheric escape, with the proximity of some planets to their host stars leading to hydrodynamic outflows of their hydrogen-dominated atmospheres. This mass loss is capable of stripping some young sub-Neptunes of their primordial envelope to form the population of super-Earths \citep[e.g.][]{Owen2013,LopezFortney2013,Ginzburg2018,Gupta2019}. This theory is supported by observations of escaping hydrogen/helium exospheres from sub-Neptunes \citep[e.g.][]{DosSantos2023a,Loyd2025} as well as bulk density measurements which show super-Earths to be consistent with bare, rocky planets and sub-Neptunes to be consistent with hosting hydrogen-dominated envelopes \citep[e.g.][]{Weiss2014,JontofHutter2014}. 

Planet evolution, particularly in the absence of atmospheric escape, is convergent: different initial conditions can evolve to the same state after billions of years. A planet with an initially higher entropy cools and contracts more rapidly, meaning it reaches the same thermodynamic state as a planet that started with lower entropy. This fact makes inferring the initial properties of mature planets challenging. However, young planets allow these degeneracies to be broken. Several observational studies have shown that young super-Earths or sub-Neptunes with ages $\lesssim 100$~Myrs are larger than their evolved counterparts, in some cases similar to the size of gas giants ($\sim 10 R_\oplus$) although with substantially lower masses \citep[e.g.][]{Fernandes2022,Christiansen2023,Vach2024a,Barat2024,Thao2024,Dai2025,Livingston2026}. Since these planets are still hot from formation, one can infer their initial entropy \citep{Owen2020} or their envelope mean molecular weight \citep{Rogers2025b}. 

In this study, we utilise young exoplanets to test the prevalence of hydrogen-silicate miscibility in sub-Neptunes. Recent studies have shown that hydrogen and silicate melt can become miscible at the thermodynamic temperatures ($\gtrsim 4000$~K) and pressures ($\sim$~GPa) relevant to the interiors of sub-Neptunes,  \cite[e.g.][]{Young2024,Gilmore2026}. This implies that hydrogen and silicate combine to form a homogeneous super-critical miscible interior that exsolves (releases) hydrogen into the planet's immiscible envelope over time \citep{Rogers2025c}. This is contrary to previous frameworks which have assumed discrete planetary layers. Furthermore, miscibility is a generalisation of solubility \citep[as adopted in multiple studies e.g.][]{Chachan2018,Kite2019,Kite2021,Schlichting2022}. Solubility as an assumption breaks down at high temperatures and high solute concentrations, which is precisely the region of parameter space relevant for the deep interiors of sub-Neptunes. \citet{Rogers2025c} showed that miscibility affects the thermal evolution of sub-Neptunes, with significant implications for young planets. Here, we couple an evolving interior structure framework for miscible sub-Neptunes with atmospheric escape models to predict how and where to find such planets in the young planet population.

\section{Method} \label{sec:method}
We simulate sub-Neptune evolution with an interior structure model based on that of \citet{Rogers2025c}. We provide a brief summary here, in addition to details on model upgrades, but point the interested reader to this latter study for more information.

\subsection{Interior structure model} \label{sec:interior_structure_method}
The interior structure framework of \citet{Rogers2025c} assumes spherically-symmetric $1$-D planetary models in hydrostatic and thermochemical equilibrium. It solves the following equations for radial, pressure and temperature structure:
\begin{equation} \label{eq:mass_cons}
    \frac{\partial r}{\partial m} = \frac{1}{4 \pi r^2 \rho},
\end{equation}

\begin{equation} \label{eq:hydro_eq}
    \frac{\partial P}{\partial m} = -\frac{Gm}{4 \pi r^4},
\end{equation}

\begin{equation} \label{eq:heat_transport}
    \frac{\partial T}{\partial m} = -\frac{Gm}{4 \pi r^4} \frac{T}{P} \nabla,
\end{equation} 
where $m$ is the mass contained within radius $r$, $\rho$ is the density, $P$ is the pressure, $T$ is the temperature, $G$ is the gravitational constant and $\nabla \equiv \partial \ln T / \partial \ln P$ is the temperature gradient and depends on the material properties. In practice, these equations are solved using an adaptive step \verb|RK45| shooting method with an outer boundary condition such that $P(m=M_\text{p})=0.1$~bar and $T(m=M_\text{p})=T_\text{eq}$, where $M_\text{p}$ is the planet mass and $T_\text{eq}$ is the planet equilibrium temperature (assuming zero Bond albedo), which are both free parameters of the model.

To account for hydrogen-silicate miscibility, the model utilises the binodal surface derived from the H$_2$-MgSiO$_3$ density functional theory (DFT) molecular dynamics simulations from \citet{Gilmore2026}. In regions of the planet with temperatures and pressures greater than the binodal surface, hydrogen and silicate melt are miscible and combine to form a supercritical fluid. At temperatures and pressures below the binodal surface, hydrogen and silicate melt exist in two distinct phases: gas, such as H$_2$, SiO, etc., and melt, such as MgSiO$_3$ melt with dissolved H$_2$. We highlight again that a miscibility framework is a generalisation of a solubility law, which breaks down at high temperatures and high solute concentration. While a non-ideal mixing framework, as used in this study, correctly predicts the existence of a binodal surface, a solubility law does not.

To account for the mixing of hydrogen and silicate in the miscible interior, we follow the best-fit to DFT results from \citet{Young2024}, in which the density of the uncompressed H$_2$-MgSiO$_3$ mixture is given by:

\begin{equation} \label{eq:Vmix}
    \rho_{\text{mix},0} = (x_{\text{H}_2} \mu_{\text{H}_2} + x_{\text{sil}} \mu_{\text{sil}}) \, \bigg(x_{\text{H}_2} \frac{\rho_{\text{H}_2,0}}{\mu_{\text{H}_2}} + x_{\text{sil}} \frac{\rho_{\text{sil},0}}{\mu_{\text{sil}}} \bigg),
\end{equation}
where $\rho_{\text{H}_2,0}$ and $\rho_{\text{sil},0}$ are the uncompressed densities of pure hydrogen and silicate. Unlike in \citet{Rogers2025c}, we use the \citet{Wolf2018} Vinet fit to the updated \citet{Luo2025} equation of state for MgSiO$_3$ melt. For hydrogen, we again use the equation of state from \citet{Chabrier2019}. These equation of state tables are also used to compute adiabatic temperature gradients, which are used in Equation \ref{eq:heat_transport} in convective regions of the planet. We assume the entire miscible interior is convective, facilitating a constant hydrogen concentration throughout the interior (although this concentration can vary with time). In the immiscible envelope, heat can be transported via convection, conduction and radiative diffusion. We use the moist adiabat from \citet{Graham2021} (their Equation 1) to account for condensible silicate gas species and rain-out (which we assume is instantaneous). Additionally, we account for convection inhibition due to mean molecular weight gradients following \citet{Markham2021,Markham2022}. We compute thermal conductivities from experimental results from \citet{McWilliams2016}. Finally, we assume a simple power-law relation for Solar-metallicity Rosseland mean gas opacities \citep[e.g.][]{Rogers2010,Owen2017}:
\begin{equation} \label{eq:opacity}
    \kappa = 1.3 \times 10^{-2} \bigg( \frac{T}{1000 \, \text{K}} \bigg)^{0.45} \bigg( \frac{P}{1 \, \text{bar}} \bigg)^{0.68} \text{ cm}^2 \text{ g}^{-1}.
\end{equation}

\subsection{Thermal evolution and atmospheric escape} \label{sec:evolution_method}
In order to evolve our planetary structure model, we numerically solve two ordinary differential equations for global energy and mass conservation. For energy, we solve:
\begin{equation} \label{eq:GlobalEnergyEvolve}
    \frac{d E_\text{p}}{d t} = - L_\text{rcb} + \frac{GM_\text{p}\dot{M}}{R_\text{rcb}},
\end{equation}
where the first and second terms represent energy lost due thermal cooling and mass loss, respectively. Here, $L_\text{rcb}$ is the radiative luminosity at the outer-most radiative-convective boundary of the model, located at $R_\text{rcb}$, and $E_\text{p}$ is the total energy of the planet, given by the sum of thermal and gravitational potential energies:
\begin{equation}
    E_\text{p} = \int_0^{M_\text{p}} \bigg( \bigg[\sum_i c_{\text{p},i} \,X_i(m) \,T(m) \bigg]- \frac{G\,m}{r} \bigg) \;dm,
\end{equation}
where $c_{\text{p},i}$ is the specific isobaric heat capacity for component $i$, including hydrogen and silicate in melt and vapour phases, and $X_i(m)$ is the mass fraction of this component. Then, for mass conservation, we also solve:
\begin{equation} \label{eq:GlobalMassEvolve}
    \frac{d M_\text{p}}{d t} = - \dot{M},
\end{equation}
where $\dot{M}$ is the hydrogen mass loss rate, for which we use the tabulated rates from \citet{Misener2026}. These mass loss rates are calculated with the hydrodynamic radiative transfer code \verb|aiolos| \citep{Schulik2022} and account for the transition from bolometrically-driven core-powered mass loss to X-ray/extreme ultra-violet (XUV)-driven photoevaporation. These grids are constructed and interpolated as a function of planet mass, transit radius and XUV luminosity. At early times, when a planet is highly inflated and undergoing bolometrically-driven mass loss, the escape rate can exceed that which is allowed by the transport of material from the radiative-convective boundary to the base of the wind, which we approximate to be the photosphere at $R_\text{ph}$. In other words, the hydrodynamic wind can only be resupplied from the optically thick envelope with transport powered by the planet's own internal luminosity \citep{Tang2024}. This limit is given by:
\begin{equation}
    \dot{M} = \frac{L_\text{rcb}}{GM_\text{p}/R_\text{rcb} - GM_\text{p}/R_\text{ph}}.
\end{equation} 
We find that this limit is only active for the first $\sim 10^4-10^5$~yrs of evolution for typical young sub-Neptunes. For the XUV luminosity, we follow \citet{Rogers2021b} and assume the following high-energy stellar evolution:
\begin{equation}
\label{eq:LxLbol}
\begin{split}
\frac{L_\text{XUV}}{L_\text{bol}}
& =\left( \frac{L_\text{XUV}}{L_\text{bol}} \right)_{\text{sat}}
  \left( \frac{M_*}{M_\odot} \right)^{-0.5} \\
 & \times \begin{cases}
    1 & \text{if } t < t_\text{sat}, \\
    \left( \dfrac{t}{t_\text{sat}} \right)^{-1-a_0}, & \text{if } t \ge t_\text{sat}
  \end{cases}
\end{split}  
\end{equation}

where $a_0=0.5$ and $(L_\text{XUV} / L_\text{bol})_\text{sat} = 10^{-3.5}$ \citep{Wright2011, Jackson2012}. The saturation time $t_\text{sat}$ follows:
\begin{equation} \label{eq:tsat}
    t_\text{sat} = 10^{2} \; \bigg( \frac{M_*}{M_\odot} \bigg)^{-1.0} \;\; \text{Myr},
\end{equation}
which is motivated by observational studies \citep{Jackson2012,Shkolnik2014,McDonald2019}. 

\begin{figure*}
	\includegraphics[width=2.0\columnwidth]{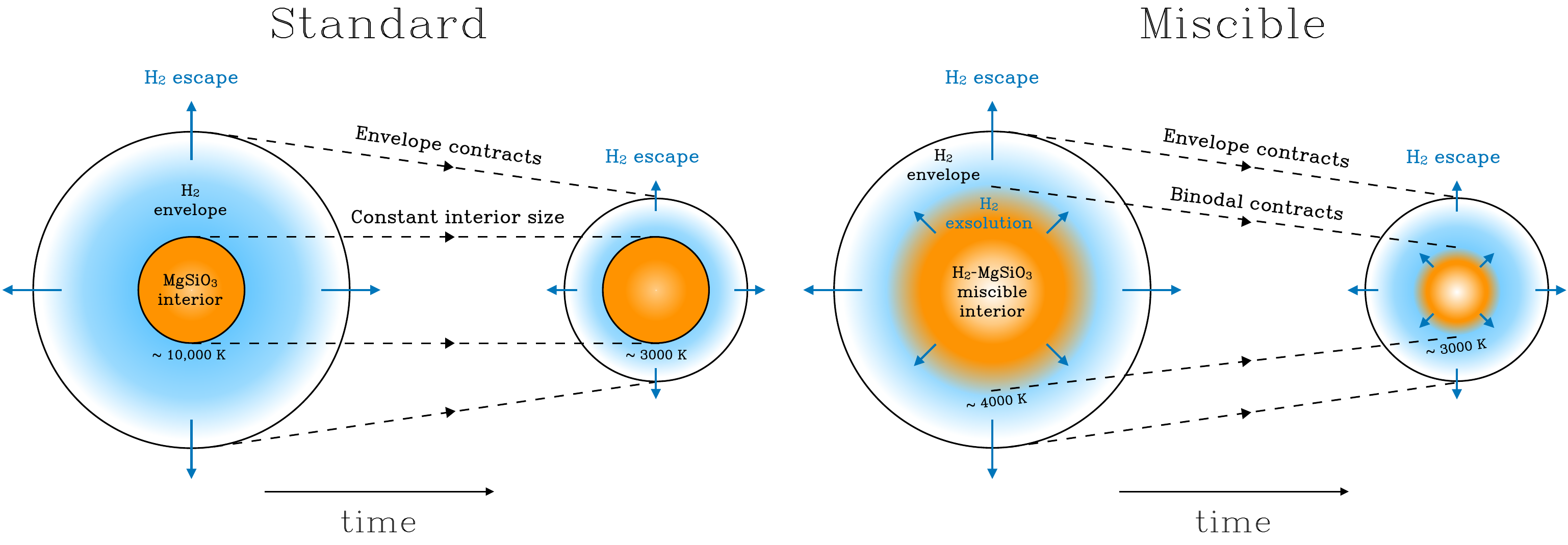}
    \centering
        \cprotect\caption{Schematic showing the thermal and atmospheric escape evolution of two interpretations of sub-Neptune interiors. In a standard planetary model, a silicate interior is distinct from a hydrogen-rich envelope. As the planet contracts, hydrogen is removed from the upper atmosphere via stellar heating, and the interior-envelope boundary does not significantly contract with time. The temperature at this interface can reach of order $\sim 10,000$~K at early ages. In the miscible model, the interior-envelope boundary is defined by a \textit{binodal surface}, which delineates regions in which hydrogen and silicate are miscible or immiscible. The temperature at the binodal surface does not change significantly with time. The radial position of the binodal surface contracts with time, exsolving hydrogen into the base of the envelope, while hydrogen is also removed from the top of the upper atmosphere.} \label{fig:Schematic_evolution} 
\end{figure*}

To evolve the equations for global energy and mass conservation (Equations \ref{eq:GlobalEnergyEvolve} and \ref{eq:GlobalMassEvolve}), we produce grids of models at various planet energies, $E_\text{p}$, silicate masses, $M_\text{sil}$, and hydrogen mass fractions, $f_{\text{H}} \equiv M_\text{H} / M_\text{p}$, where $M_\text{H}$ is the total hydrogen mass across the miscible interior and immiscible envelope. We then interpolate across this grid, while enforcing energy and mass conservation according to Equations \ref{eq:GlobalEnergyEvolve} and \ref{eq:GlobalMassEvolve} with a $1^\text{st}$-order Euler method. In order to maintain numerical stability, we ensure the timestep $\Delta t \ll \text{min} (t_\text{cool}, t_{\dot{M}})$, where $t_\text{cool} = |U| / L_\text{rcb}$ is the cooling timescale (analogous to the Kelvin-Helmholtz timescale) and $U$ is the total gravitational potential energy of the planet, and $t_{\dot{M}} = M_\text{H} / \dot{M}$ is the hydrogen mass loss timescale. At each timestep, we attach an isothermal atmosphere above the structure model in order to locate the transit radius at a pressure of $50$~mbar \citep[e.g.][]{Lopez2014,Misener2026}.

We begin each simulation after disc dispersal has concluded, and prescribe each planet's initial entropy with a cooling timescale. To begin, we assume a fiducial case of $t_\text{cool} = 100$~Myrs, motivated by simulations of boil-off \citep{Owen2016, Rogers2024a}. We then remove this assumption in Section \ref{sec:test_definitions}.

\section{Results} \label{sec:results}

We begin by comparing two end-member models: one in which sub-Neptune interiors are fully miscible, as previously described in Section \ref{sec:method}, and another in which miscibility does not operate in sub-Neptune interiors. We label this latter end-member as a ``standard'' model. In this case, a planet's interior and envelope are physically and chemically distinct, with the interior consisting of pure silicate melt and the envelope of pure hydrogen. We note that, in reality, scenarios may exist in which sub-Neptunes are only partially miscible, which we discuss in Section \ref{sec:uncertainties}.


\subsection{A balance between hydrogen exsolution and escape} \label{sec:escape_vs_exsolution}
Hydrogen-silicate miscibility allows for hydrogen to be stored in a planet's miscible interior. As a planets cools and contracts, its binodal surface (which delineates the miscible interior and immiscible envelope) moves radially inward, releasing hydrogen into the envelope \citep{Rogers2025c}. We refer to this process as hydrogen exsolution.\footnote{Silicate vapour and melt are also exsolved into the envelope as the binodal surface moves radially inwards, however, this only contributes a very small silicate mass into the envelope.} While hydrogen exsolution acts as an envelope mass source, atmospheric escape will act as an envelope mass sink, with the upper-most layers being removed by stellar-heating. This interplay is shown as a schematic in the right-hand side of Figure \ref{fig:Schematic_evolution}, with blue arrows showing the relevant sources and sinks within the envelope. In the left-hand side of Figure \ref{fig:Schematic_evolution}, we show a schematic for a standard planetary model, which only includes the envelope sink due to atmospheric escape. Since miscibility does not occur, the size of the interior remains approximately constant, as found in \citet{Rogers2025c}.

\begin{figure*}
	\includegraphics[width=2.0\columnwidth]{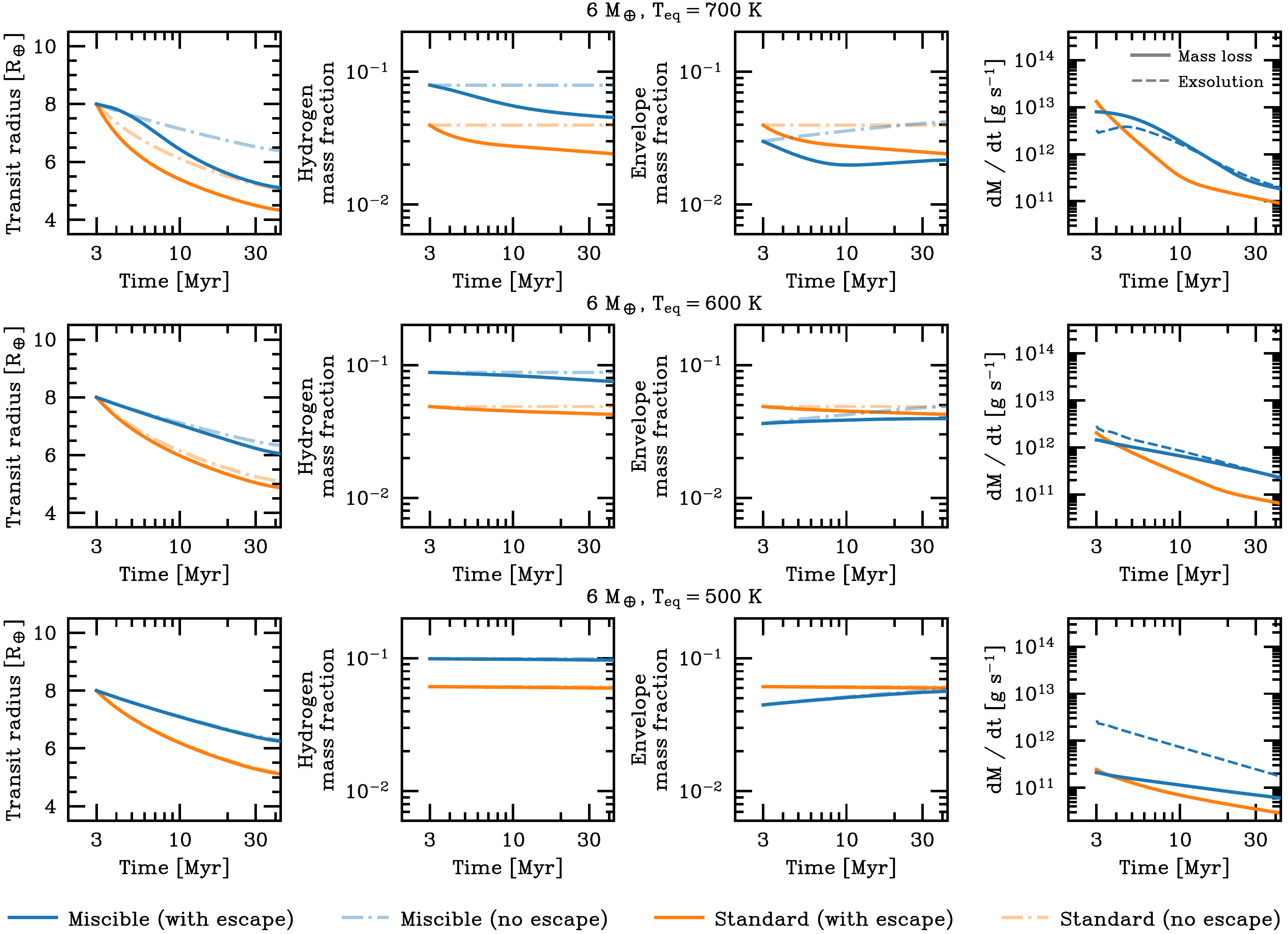}
    \centering
        \cprotect\caption{Evolution of a fiducial sub-Neptune with silicate mass of $6 M_\oplus$ and equilibrium temperatures of $700$~K, $600$~K and $500$~K in the top, middle and bottom rows, respectively. Miscible and standard models are shown in blue and orange, respectively. All models are given an initial hydrogen mass fraction required to yield an initial transit radius of $8R_\oplus$. All models begin with an initial cooling timescale of $100$~Myrs. The planet's transit radius, hydrogen mass fraction, envelope mass fraction are shown in the first, second and third columns. We show the hydrogen mass loss rate and hydrogen exsolution rate in the fourth column with solid and dotted lines, respectively. Models which do not include atmospheric escape are shown in dot-dashed lines.} \label{fig:Fiducial} 
\end{figure*}

In Figure \ref{fig:Fiducial}, we show the evolution of our fiducial sub-Neptune with silicate mass of $6M_\oplus$ and equilibrium temperatures of $700$~K, $600$~K and $500$~K orbiting a $1M_\odot$ star. We begin all simulations at $3$~Myrs, corresponding to a typical protoplanetary dispersal time \citep[e.g.][]{Kenyon1995,Ercolano2011,Koepferl2013}. In the first panel, we show the transit radius as a function of time. Miscible models are shown in blue, while standard models are shown in orange. In order for all models to begin with the same initial transit radius of $8R_\oplus$ with the same initial cooling timescale of $100$~Myrs, the miscible models require a larger hydrogen mass fraction, since much of their hydrogen content is initially stored within their interiors. In all cases, simulations that do not incorporate atmospheric escape, shown in dot-dashed lines, demonstrate significantly slower radial contraction than in the case with mass loss. With or without escape, miscible sub-Neptunes contract slower than standard models since hydrogen and silicate are redistributing within the planet over time which consumes much of their energy budget \citep{Rogers2025c}. 

In the second column of Figure \ref{fig:Fiducial}, we show the hydrogen mass fractions of our fiducial sub-Neptune. In the case with no atmospheric escape, the hydrogen mass fractions remain constant with time. Recall that our miscible model requires a higher initial hydrogen mass fraction in order to reproduce the same initial size for a given initial cooling timescale. In the case with escape, the hydrogen mass fractions decreases monotonically as stellar heating removes hydrogen from the upper atmosphere. This is true for both standard and miscible models. 

In third column of Figure \ref{fig:Fiducial}, we show the envelope mass fractions of our models. For the standard sub-Neptune, the hydrogen and envelope mass fractions are equal, since all of the hydrogen exists in the envelope. However, in the miscible model, hydrogen partitions between the interior and envelope. As found in \citet{Rogers2025c}, the case with no atmospheric escape results in envelope mass fractions that increase as hydrogen exsolves from the interior. However, with mass loss included, the envelope mass fractions can vary non-monotonically due to the balance between hydrogen escape rates and exsolution rates, which we also show in the fourth column in solid and dotted lines, respectively. For example, in the case of $T_\text{eq}=700$~K, the envelope mass fraction initially decreases, since the initial bolometric core-powered mass loss rate exceeds the hydrogen exsolution rate. After $\sim10$~Myr, the envelope mass fraction increases since the exsolution surpasses mass loss once the escape regime switches from bolometric to XUV-driven. 

\subsection{A population-level test for the prevalence of miscible sub-Neptunes} \label{sec:test_definitions}

\begin{table*}
\centering
\begin{tabular}{lcccccc}
\hline
Planet & $M_p$ ($M_\oplus$) & $R_p$ ($R_\oplus$) & $T_{\rm eq}$ (K) & Age (Myr) & $M_*$ ($M_\odot$) & Ref. \\
\hline
HIP 67522 b  & $13.8\pm1.0$ & $9.8\pm0.5$  & $1176\pm20$ & $17\pm2$  & $1.22\pm0.05$  & (a) \\
V1298 Tau c  & $4.7\pm0.6$  & $5.1\pm0.4$  & $953\pm36$  & $20\pm10$ & $1.10\pm0.05$  & (b) \\
V1298 Tau d  & $6.0\pm0.7$  & $6.5\pm0.4$  & $831\pm31$  & $20\pm10$ & $1.10\pm0.05$  & (b) \\
V1298 Tau b  & $13.1\pm5.3$ & $9.4\pm0.6$  & $666\pm25$  & $20\pm10$ & $1.10\pm0.05$  & (b) \\
V1298 Tau e  & $15.3\pm4.2$ & $10.2\pm0.8$ & $527\pm20$  & $20\pm10$ & $1.10\pm0.05$  & (b) \\
TOI-2076 c   & $7.2\pm1.4$  & $3.5\pm0.1$  & $581\pm13$  & $210\pm20$ & $0.84\pm0.02$ & (c) \\
TOI-1136 d   & $8.4\pm1.7$  & $4.6\pm0.1$  & $843\pm6$   & $700\pm150$ & $1.02\pm0.03$ & (d), (e) \\
TOI-1136 f   & $9.7\pm3.8$  & $3.7\pm0.1$  & $658\pm5$   & $700\pm150$ & $1.02\pm0.03$ & (d), (e) \\
TOI-5398 c   & $11.8\pm4.8$ & $3.5\pm0.2$  & $1242\pm37$ & $650\pm150$ & $1.15\pm0.01$ & (f) \\
\hline
\end{tabular}
\caption{Planetary parameters for our chosen sample. (a) \citet{Thao2024}; (b) \citet{Livingston2026}; (c) \citet{Wang2026}; (d) \citet{Dai2023}; (e) \citet{Polanski2024}; (f) \citet{Mantovan2024}.}
\label{tab:planets}
\end{table*}

We now aim to test the prevalence of miscible sub-Neptunes in the exoplanet population. As shown in Section \ref{sec:escape_vs_exsolution}, miscibility causes planets to contract slower than standard models with no chemistry. The magnitude to which this is true varies across a range in planet mass, equilibrium temperature and age. Therefore, in order to probe this effect, we focus on observed exoplanets with measured masses, radii and ages $\lesssim 1$~Gyr. The framework behind this test is a Bayesian hierarchical inference model. By fitting the observed planet sample with both standard and miscible frameworks, we can perform model comparison to determine if the data favour one model over the other.

In this framework, the models are required to reproduce a population of planets' transit radii and masses at their observed ages and equilibrium temperatures \citep[see also][for a similar inference setup applied to individual planets]{Owen2020}. To do so, we construct large grids of planet evolution models, spanning silicate masses of $M_\text{sil} \in [4,\,15]M_\oplus$, equilibrium temperatures of $T_\text{eq} \in [500,\,1300]$~K, initial hydrogen mass fractions of $f_\text{H,init} \in [0.005, 0.4]$ and initial cooling timescales of $t_\text{cool,init} \in [3\times10^6, 3\times10^{10}]$~yr. We interpolate across these grids to produce each evolution track in both miscible and standard frameworks. We do not model the complete envelope stripping of sub-Neptunes and resultant transition to the super-Earth regime. This requires an understanding of silicate melt crystallisation, which is beyond the scope of this study. Instead, if a planet reaches a hydrogen mass fraction of $<0.005$, we simply calculate a final transit radius based on the pure solid silicate mass-radius relations of \citet{Fortney2007}. 

For each observed planet in the sample, we include $5$ model parameters which we aim to infer: the planet's silicate mass, $M_\text{sil}$; initial hydrogen mass fraction, $f_\text{H,init}$; initial cooling timescale at disc dispersal, $t_\text{cool,init}$; present-day equilibrium temperature, $T_\text{eq}$; and the time since disc dispersal, $t_\text{evolve}$. We place uniform priors on each silicate mass and log-uniform priors on each initial hydrogen mass fraction and initial cooling timescale. We place Gaussian priors on each equilibrium temperature, with mean and standard deviation according to the present-day observed values. The time since disc dispersal, $t_\text{evolve}$, accounts for the uncertainty in both stellar age and the time at which the protoplanetary disc dispersed. We follow \citet{Owen2020} and place the following prior on $t_\text{evolve}$:
\begin{equation} \label{eq:dispersal_prior}
    \begin{split}
    P(t_\text{evolve}) & = \frac{1}{2\sigma_d} \, \exp \bigg \{ \frac{\sigma_*^2 + 2 \sigma_d (t_\text{evolve} + t_d - t_*)}{2\sigma_d^2} \bigg \} \\
    & \times \bigg \{ 1- \text{erf} \bigg[\frac{\sigma_*^2 + \sigma_d (t_\text{evolve} + t_d - t_*)}{\sqrt{2} \sigma_* \sigma_d}  \bigg] \bigg \}.
    \end{split}
\end{equation}
This prior distribution accounts for the exponential fall-off in disc occurrence with $t_d = 1$~Myr and $\sigma_d=3$~Myr \citep[e.g.][]{Mamajek2009}. For each planet in the observed sample, we assume its host star has an age of $t_*$ with a Gaussian uncertainty of $\sigma_*$. In the inference model, each observed planet in a multi-planet system shares the same parameter for time since disc dispersal. 

We also include a set of hyper-parameters which inform the overall distribution of initial hydrogen mass fractions and initial cooling timescales across the population of observed planets. The initial hydrogen mass fraction distribution is a joint-power law with silicate mass and equilibrium temperature:
\begin{equation} \label{eq:fH_init}
    f_\text{H,init} = f_0 \, \bigg( \frac{M_\text{sil}}{5 M_\oplus}\bigg)^\alpha \, \bigg( \frac{T_\text{eq}}{1000 \text{ K}}\bigg)^\beta,
\end{equation}
where we seek to infer the hyper-parameters $f_0$, $\alpha$ and $\beta$. Then, the initial cooling timescales are drawn from a log-Gaussian distribution:
\begin{equation} \label{eq:tcool_init}
    \log_{10} (t_\text{cool,init} / \text{Myr}) \sim \mathcal{G}(\mu_t, \sigma_t),
\end{equation}
where we seek to infer the hyper parameters $\mu_t$ and $\sigma_t$. We place uniform priors on $\alpha$, $\beta$, $\mu_t$ and $\sigma_t$, and log-uniform priors on $f_0$. {\jr{Priors on $f_\text{H,init}$ and $t_\text{cool,init}$ are truncated to the edges of the model grid.}}
\newpage
{\jr{The likelihood function is then given as:

\begin{equation} \label{eq:likelihood}
    \begin{split}
        & \ln \mathcal L  = \sum_i -\frac{1}{2} \bigg[  \bigg( \frac{R_i - R_i^{\text{obs}}}{\sigma_{R,i}} \bigg)^2  + \bigg( \frac{M_i - M_i^{\text{obs}}}{\sigma_{M,i}} \bigg)^2\bigg] \\
        & + \sum_i \bigg[ -\frac{1}{2} \bigg( \frac{\log_{10} f_{\text{H,init},i} - \mu_{f,i}}{\sigma_f} \bigg)^2  - \ln(A_{f,i} \, \sigma_f \sqrt{2\pi} ) \bigg] \\
        & + \sum_i \bigg[ -\frac{1}{2} \bigg( \frac{\log_{10} t_{\text{cool,init},i} - \mu_{t}}{\sigma_t} \bigg)^2  - \ln(A_{t,i} \, \sigma_t \sqrt{2\pi} ) \bigg], 
    \end{split}
\end{equation}
where $\sigma_f$ is the final hyper-parameter to infer, and controls the astrophysical scatter away from the joint-power law for initial hydrogen mass fraction in Equation \ref{eq:fH_init}. Here, $A_{f,i}$ and $A_{t,i}$ are normalisation factors that account for truncation of the Gaussian likelihood factors at the edge of the evolution grids.

}}

\subsubsection{Planet sample}

As a test case of our inference model, we apply it to a sample of $9$ planets with ages $< 1$~Gyr. These planets were chosen based on their mass uncertainties $\lesssim 40 \%$, radius uncertainties $\lesssim 10\%$ and age uncertainties $\lesssim 40 \%$. Namely, we include V1298 Tau b, c, d and e \citep{David2019a,David2019b,Livingston2026}, HIP 67522 b \citep{Barber2024b,Thao2024}, TOI-2076 c \citep{Hedges2021,Wang2026}, TOI-1136 d and f \citep{Dai2023,Polanski2024}, and TOI-5389 c \citep{Mantovan2024}. Each planet's mass, transit radius, equilibrium temperature, age and stellar mass are shown in Table \ref{tab:planets}. 

In general, this sample can be split into two groups: a young subset (V1298 Tau b, c, d, e, HIP 67522 b, with ages $< 0.1$~Gyr) and and adolescent subset (TOI-2076 c, TOI-1136 d, f and TOI-5389 c with ages from $0.1-1$~Gyr). We see that the young group is more inflated, with radii between $\approx 5-10 \, R_\oplus$, while the adolescent group is more contracted, with radii $\lesssim 5\,R_\oplus$. This is indicative of thermal contraction and atmospheric escape on $\sim 100$~Myr timescales. While the sample is small in size, it spans a wide range in planet mass, radius, equilibrium temperature and age, which is beneficial for the inference model to discern which of the standard and miscible models is statistically preferred.

With a sample of $9$ planets, the inference model requires a total of 47 parameters. We numerically evaluate the posterior distribution for both standard and miscible models using \verb|ultranest| \citep{Buchner2021}. Both chains were run to effective sample sizes of $N_\text{eff} \approx 10^4$.

\subsubsection{Inference model results} \label{sec:MCMCresults}

\begin{figure*}
	\includegraphics[width=2.0\columnwidth]{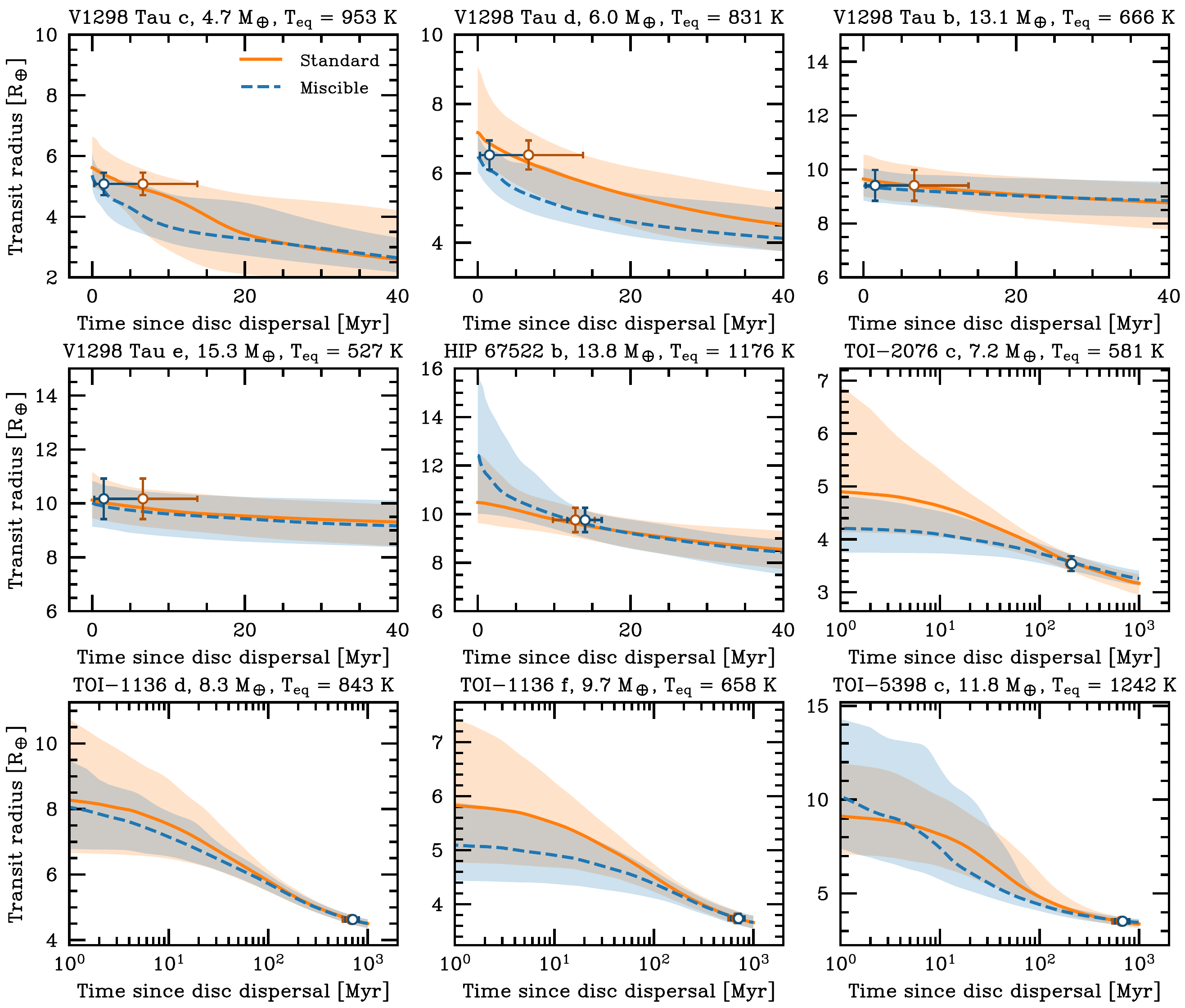}
    \centering
        \cprotect\caption{The predicted evolution in transit radius for our planet sample, with standard and miscible model fits shown in orange and blue, respectively. Shaded regions represent the $1\sigma$ uncertainties. Each planet's observed transit radius at its inferred time since disc dispersal is shown as a black circle.} \label{fig:Evolution} 
\end{figure*}

We present the best-fit evolution models for the $9$ planets in our sample in Figure \ref{fig:Evolution}. Standard and miscible models are shown in orange and blue, respectively, with $1\sigma$ uncertainties shown in shaded regions. All planet properties are successfully reproduced at their observed ages under both frameworks. The slower radial contraction in miscible model from Figure \ref{fig:Evolution} is less prominent here since we must also infer each planet's initial cooling timescale which controls the initial size and contraction rate. We present the inferred initial hydrogen mass fractions and cooling timescales in Appendix \ref{app:posteriors}. For an individual planet, \citet{Owen2020} showed that one can only infer a lower limit on an initial cooling timescale since above some value there is always an longer initial cooling timescale that can balance an increased initial envelope mass to match a planet's size. We highlight, however, that the statistical power of this inference problem is due to the fact that it operates at a population level. As a result, some of these degeneracies can be broken due to the requirement of describing the entire sample (while still accounting for astrophysical scatter within the underlying distributions). The inferred hyperparameters that control population-level distributions for initial hydrogen mass fraction and cooling timescale are also shown in Appendix \ref{app:posteriors}. We find typical initial cooling timescales of $\gtrsim 0.1-1$~Gyr for both models, which is indicative of dramatic boil-off processes during disc dispersal which prematurely cool a planet \citep[e.g.][]{Owen2016,Rogers2024a}. We also find initial hydrogen mass fractions to scale approximately as $f_\text{H,init} \propto M_\text{sil}^{2}$, which is also consistent with boil-off \citep{Rogers2024a}.

The benefit of Bayesian inference is the ability to perform model comparison. In our case, for which both models share the same set of parameters and prior probabilities, the Bayes factor, $\mathcal{K}$, is the ratio of marginalised posterior probabilities under each model:
\begin{equation} \label{eq:BayesFactor}
    \mathcal{K} = \frac{P( \,\text{miscible} \,| \, \mathcal{D})}{P(\,\text{standard} \, | \, \mathcal{D})},
\end{equation}
where $\mathcal{D}$ is the data. {\jr{We find a Bayes factor of $\mathcal{K} = 3.48^{+5.11}_{-2.07}$, which represents mild preference in favour of the miscible model, albeit with large uncertainties due to Poisson noise. The allowed values at the lower $1\sigma$ limit of $\mathcal{K} \approx 1.41$ represent no statistical preference towards either model.}} Note that strong evidence in favour of the fully miscible model would require a Bayes factor of $\mathcal{K}\gtrsim20$ \citep{Kass1995}. In our case, this lack of evidence is primarily due to a small planet sample size.

\subsection{How many planets detections are needed?} \label{sec:forecasting}

As shown in Section \ref{sec:test_definitions}, we currently have an insufficient planet sample to determine the statistical preference for fully miscible sub-Neptunes. We now ask the question: how large a sample is needed to find strong evidence in favour of one model over the other? To do so, we run mock planet surveys of varying size to find the threshold above which strong evidence is a likely outcome. For each mock survey of a given model type, we randomly draw silicate masses from the Bernstein polynomial distribution inferred by \citet{Rogers2021} (their Model II), which peaks at $\sim 4M_\oplus$. We also draw random orbital periods from the following distribution:
\begin{equation} \label{eq:Porb_dist}
    \frac{dN}{dP} \propto \frac{1}{(\frac{P}{P_0})^{-k_1} + (\frac{P}{P_0})^{-k_2}},
\end{equation}
where $P_0 = 5.5$~days, $k_1 = 2.3$, and $k_2=0.0$ as found in \citet{Rogers2021} when fit to evolved \textit{Kepler} planets. Here we implicitly assume that the orbital period distribution does not change with time, for example due to post-disc orbital migration. We draw random host stellar masses from a Gaussian distribution with a mean of $1M_\odot$ and standard deviation of $0.2M_\odot$ and assume a stellar mass-luminosity relation for a magnitude-limited transit survey of $L_* = L_\odot (M_* / M_\odot)^{4.5}$ \citep[e.g.][]{Rogers2021b}. For each survey, we draw hyperparameters from the \verb|ultranest| posterior distributions from Section \ref{sec:test_definitions} and then assign each planet an initial hydrogen mass fraction and initial cooling timescale according to Equations \ref{eq:fH_init} and \ref{eq:tcool_init}. We randomly draw protoplanetary disc dispersal times from the exponential decay distribution used to construct Equation \ref{eq:dispersal_prior}, informed from disc population studies \citep[e.g.][]{Kenyon1995,Ercolano2011,Koepferl2013}. For simplicity, we do not account for the variability in stellar pre-main sequence bolometric luminosity, meaning that the equilibrium temperature remains constant in our planet simulations. 

\begin{figure*}
	\includegraphics[width=2.0\columnwidth]{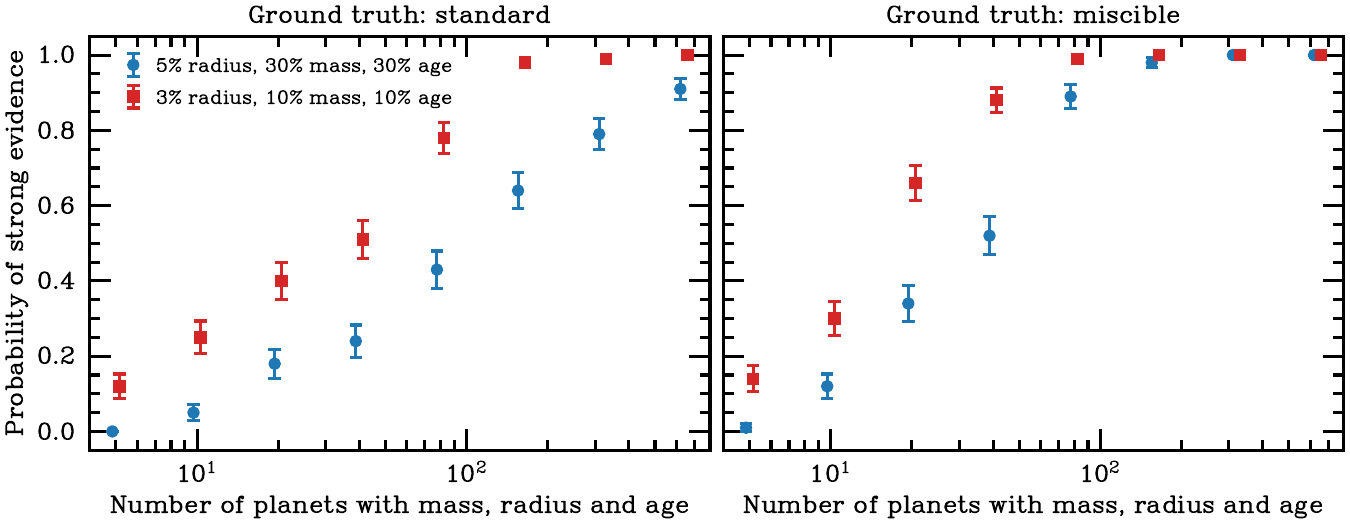}
    \centering
        \cprotect\caption{Predictions for the probability in finding strong statistical evidence (Bayes factor $\mathcal{K} > 20$, see Equation \ref{eq:BayesFactor}) in favour of standard or miscible models. In the left-hand panel, we assume standard models are the ground truth, while miscible models are assumed as the ground truth in the right-hand panel. The fraction of surveys that correctly identify the ground truth are shown for various survey sizes and two choices in measurement uncertainty as red squares and blue circles.} \label{fig:Forecast} 
\end{figure*} 

Finally, we also apply the strong biases and measurement uncertainty associated with planet detections. We use the completeness model from \citet{Rogers2025b} for a \textit{TESS}-like survey of young planets based on the analyses of \citet{Vach2024a,Fernandes2025}. This introduces strong bias against the detection of planets with orbital periods $\gtrsim 30$~days and transit radii $\lesssim 4R_\oplus$. For each randomly drawn planet, we calculate its evolution from disc dispersal time to an ``observation'' time, drawn log-uniformly between $3$~Myr and $1$~Gyr. If the planet's associated probability of being detected, which accounts for the transit probability and signal-to-noise, is sufficiently high, the planet is ``observed''. If the observation time is before the disc dispersal time, the planet is discarded from the sample. We point the interested reader to \citet{Rogers2025b} for more details of this completeness model. If a given planet is observed within our population model, we introduce one of two options for uncertainty in planet mass, transit radius and age. The first option is conservative, with Gaussian uncertainties of $30\%$, {\jr{$5\%$}}, $30\%$, respectively. The second option is optimistic, with Gaussian uncertainties of $10\%$, $3\%$, $10\%$, respectively

Each mock survey has an underlying ground truth: either standard or miscible interior structure models, in addition to a set of population-level hyperparameters. For each survey, we numerically solve for the maximum-likelihood under both standard and miscible models using the full hierarchical inference model presented in Section \ref{sec:test_definitions}. {\jr{The ratio of these likelihoods provides an estimate to the Bayes factor in Equation \ref{eq:BayesFactor}.}} If the Bayes factor is $\mathcal{K} > 20$ in favour of the true underlying model, then we consider this as strong statistical evidence. 

For each survey size, we perform $100$ mock surveys under both standard and miscible ground truths. The fraction of these surveys that favour the correct ground truth with strong evidence (Bayes factor $\mathcal{K} > 20$) provides an estimate for the probability of correctly identifying the preferred underlying model. We show this in \ref{fig:Forecast}, with left and right-hand panels showing standard and miscible ground truths, respectively. Uncertainties in this probability are given by a standard binomial distribution error of $\sigma = \sqrt{p(1-p) / N}$, where $p$ is the probability of strong evidence and $N=100$ is the number of trials. {\jr{The maximum-likelihood ratio method is not performed with a full Monte Carlo Markov Chain (MCMC) for the sake of computational cost. As a result, it does not estimate the uncertainty in the Bayes factor for each mock survey. However, we find that the Poisson noise from resampling the surveys is comparable to a full MCMC Bayes factor uncertainty. We confirm this by performing our nested sampling analysis from Section \ref{sec:test_definitions} for a small subset of the mock surveys. For all surveys, the maximum likelihood Bayes factor is consistent with that derived from \verb|ultranest| and the variance in Bayes factor from different survey realisations is comparable to the uncertainty in the Bayes factor itself.}}

{\jr{For conservative measurement uncertainties (blue circles), we find that strong evidence is found in $90\%$ of surveys with $N\gtrsim600$ and $N\gtrsim100$ for standard and miscible ground truths, respectively. For optimistic measurement uncertainties (red squares), we find these numbers are reduced, with strong evidence found in $90\%$ of surveys with $N\gtrsim100$ and $N\gtrsim60$ planets for standard and miscible ground truths. We highlight that these estimates are intentionally conservative given that we are sampling across ground truth hyperparameters from the existing MCMC from Section \ref{sec:MCMCresults}. This causes the disparity in survey requirement between standard and miscible ground truths for conservative measurement uncertainty due to their each model's MCMC variance. With this in mind, we summarise that a survey of $\sim 100-200$ planets will provide a $\gtrsim 90\%$ probability of yielding strong statistical evidence in favour of the correct model. 
}}

We highlight that, while we have applied observational biases to the mock surveys, we have not performed mock completeness corrections in their analysis. Completeness corrections require homogenous surveys and a quantified understanding of how bias and false positives affect the observations \citep[e.g.][]{Christiansen2015,Fulton2017,Fulton2021,Petigura2022,Dattilo2023}. However, in our proposed survey, we require measured radii \textit{and} masses which typically come from different observational strategies (e.g. transits, transit timing variations and radial velocity). To date, no study has quantified the combined observational bias involved in these measurements for low-mass planets, and so we conservatively do not correct for them in our mock surveys. {\jr{To understand how this affects the ability to differentiate between models, we repeat the analysis, however, we now normalise the likelihood function by the detectable fraction of the population given the hyperparameters \citep[e.g.][]{Foreman-Mackey2014,Rogers2021}. Note that this marginalises over the adopted core mass distribution, which in reality is unknown in this problem. As a result, this provides only an estimate to the impact of completeness corrections.}} Nevertheless, we reassuringly find that this has negligible impact on the resultant Bayes factors and thus ability to find strong evidence in favour of the ground truth structure model. Instead, completeness corrections affect the ability to recover the ground truth in hyperparameters $(f_0, \alpha, \beta, \sigma_f, \mu_t, \sigma_t)$ used for distributions of initial hydrogen mass fraction and cooling timescales from Equation \ref{eq:fH_init} and \ref{eq:tcool_init}. With completeness corrections, we recover the ground truth for all hyperparameters for survey sizes $N\gtrsim 100$. However, without completeness corrections, some hyperparameters are never correctly recovered, most notably those controlling the initial cooling timescales $(\mu_t, \sigma_t)$.

\section{Discussion} \label{sec:discussion}

We have coupled models of atmospheric escape with interior structure that accounts for hydrogen-silicate miscibility. We have investigated the behaviour of such planets at young stellar ages and constructed a population-level test for the prevalence of miscible sub-Neptunes. Here we discuss the implications of our study.

\subsection{An observational road forward} \label{sec:road_forward}
In Section \ref{sec:test_definitions}, we devised a population-level test for the prevalence of fully-miscible sub-Neptunes which primarily exploits the difference in planet contraction under standard and miscible models. Since contraction rate is the key differentiator, we highlight that the test benefits from sufficient coverage in age up to $\sim 1$~Gyr. While the current observed exoplanet population is insufficient in number to utilise this test, future missions will help increase this sample. For the adolescent planet population ($0.1-1$~Gyr), \textit{PLATO} which will target several young clusters and associations in its long pointing fields \citep{Nascimbeni2025}. For the eldest of these stars, age measurements with uncertainties $\sim 10\%$ are possible with asteroseismology \citep{Betrisey2023}. In addition, \textit{PLATO}'s dedicated radial velocity follow-up program will be able to measure the masses for a subset of these planets. These planets will be crucial for our Bayesian inference model, however, there is currently no future plan to provide completeness products as part of the \textit{PLATO} data releases. This inhibits the recovery of population-level initial conditions, as shown in Section \ref{sec:forecasting}. We encourage biases to be quantified for both transits and radial velocity surveys to maximise the scientific output in these demographic studies.

For the youngest population of planets with ages $\lesssim 0.1$~Gyr, the proposed Early eVolution Explorer (\textit{EVE}) space telescope is particularly promising for our test of hydrogen-silicate miscibility. This mission would perform a homogenous transit survey of young stellar clusters, with a predicted yield of $\sim 100$ planets \citep{Zhou2026}. Radial velocity mass measurements are notoriously difficult for the youngest, most active stars. However, there are promising avenues for mass determination through a planet's atmospheric scale height with \textit{JWST} \citep[e.g.][]{Thao2024,Barat2024}, or transit-timing variations \citep[e.g.][]{Livingston2026}. While we have assumed all sub-Neptunes are silicate interiors with hydrogen-dominated envelopes, EVE will also determine the fraction of ``water worlds'' within the sub-Neptune population, which is not possible with planets orbiting evolved host stars with ages $\gtrsim 1$~Gyr \citep{Rogers2025b,Zhou2026}.

\subsection{Modelling limitations and uncertainties} \label{sec:uncertainties}
Our models for miscible sub-Neptunes include several assumptions that should be addressed in future work. Firstly, our interior structure models do not include other chemical species, such as iron, helium and water within the planet. Helium and water will act to increase the density of the envelope, while iron will act to increase the density of the interior. Various combinations of these species will likely become miscible at their own critical temperatures \citep{Young2025,Gupta2024,Young2026}, further affecting planetary structure and evolution. The presence of these other species will also likely affect the position of the hydrogen-silicate binodal itself. The inclusion of addition species in the model will require the calculation of additional binodal surfaces using DFT molecular dynamic simulations \citep[e.g.][]{Gupta2024,Gilmore2026}. Equations of state are also required as a function of temperature, pressure \textit{and} chemical concentration. In our case, we assume a mixture of hydrogen and silicate melt equations of state, informed from DFT molecular dynamic calculations \citep{Young2024}.

In addition, we have also implicitly assumed that the entire of the miscible interior is convective with the use of an adiabatic temperature profile. This allows for hydrogen to be mixed at a constant concentration within the interior. We have made this assumption in order to construct an end-member scenario model in which full miscibility is possible, allowing for comparison with our standard model in which no miscibility occurs. However, reality likely sits somewhere in-between these models. We speculate that the presence of hydrogen will affect the buoyancy involved in convective fluid flow, potentially reducing its efficiency in mixing hydrogen throughout the planet. While investigating partial miscibility is beyond the scope of this study, we highlight that it would likely bring the predictions of our standard and miscible models closer to together. This would require a larger survey size in order to statistically distinguish using the methods presented in Section \ref{sec:forecasting}.

\section{Conclusions} \label{sec:conclusion}

In this study we have investigated the interplay between hydrogen-silicate miscibility and atmospheric escape for sub-Neptunes in their first $\sim 1$~Gyr of evolution. We have constructed a population-level Bayesian inference model to investigate how the prevalence of miscibility might be tested with observations. Our conclusions are as follows:

\begin{itemize}
    \item Young sub-Neptunes will store a significant fraction of their hydrogen content in their interiors if hydrogen-silicate miscibility occurs. This means less hydrogen is susceptible to atmospheric escape for highly irradiated sub-Neptunes. Over time, hydrogen is exsolved from the miscible interior into the immiscible envelope. This means envelope masses may increase or decrease with time, depending on the balance between escape and exsolution.
    \vspace{0.2cm}
        
    \item Hydrogen-silicate miscibility slows the contraction of sub-Neptunes. Miscibility causes silicate mass to sit higher in the gravitational potential well due to its mixing with hydrogen. As a miscible planet loses energy, the silicate must settle deeper into the well in order for the planet to contract, which dominates the energy budget \citep[see][]{Rogers2025c}. This result stands regardless of how strong atmospheric escape is acting on the planet.
    \vspace{0.2cm}
    
    \item The slower contraction of miscible sub-Neptunes can be exploited to determine the prevalence of miscibility at the population level. We construct a hierarchical Bayesian inference model in which model comparison is used to determine the statistical preference for miscibility. For the current population of $9$ planets with measured radii, masses and ages $<1$~Gyr, both standard and miscible models reproduce the data, however we find no preference in favour of one model over the other.

    \item For samples of $N\approx100-200$ planets, we estimate a $90\%$ probability of finding strong evidence (Bayes factors $\gtrsim 20$) in favour of either standard or miscible models. This required survey size decreases with improved accuracy in planetary radii, masses and ages. However, more planets would likely be needed if partial miscibility occurs in sub-Neptune interiors. The distribution of initial conditions, including initial hydrogen mass fractions and cooling timescales, can also be inferred if completeness corrections are performed in the analysis.
\end{itemize}

This work highlights the need for future and proposed missions, such as \textit{PLATO} and \textit{EVE} to focus on detecting young planets with ages $\lesssim 1$~Gyr \citep{Zhou2026}.

\section*{Acknowledgements}
We gratefully thank the referee for their valuable comments that helped improve this paper. We also thank William Misener for sharing the mass loss rates used in this study. JGR gratefully acknowledges support from the Kavli Foundation. HES gratefully acknowledges support from NASA's Exoplanet Research Program under grant 80NSSC25K7143, and thanks the Institute of Astronomy at the University of Cambridge, for hosting her as part of their distinguished visitor series.

\newpage
\bibliography{references}{}
\bibliographystyle{aasjournal}

\appendix

\section{Inference model posteriors} \label{app:posteriors}
\begin{figure*}
	\includegraphics[width=1.0\columnwidth]{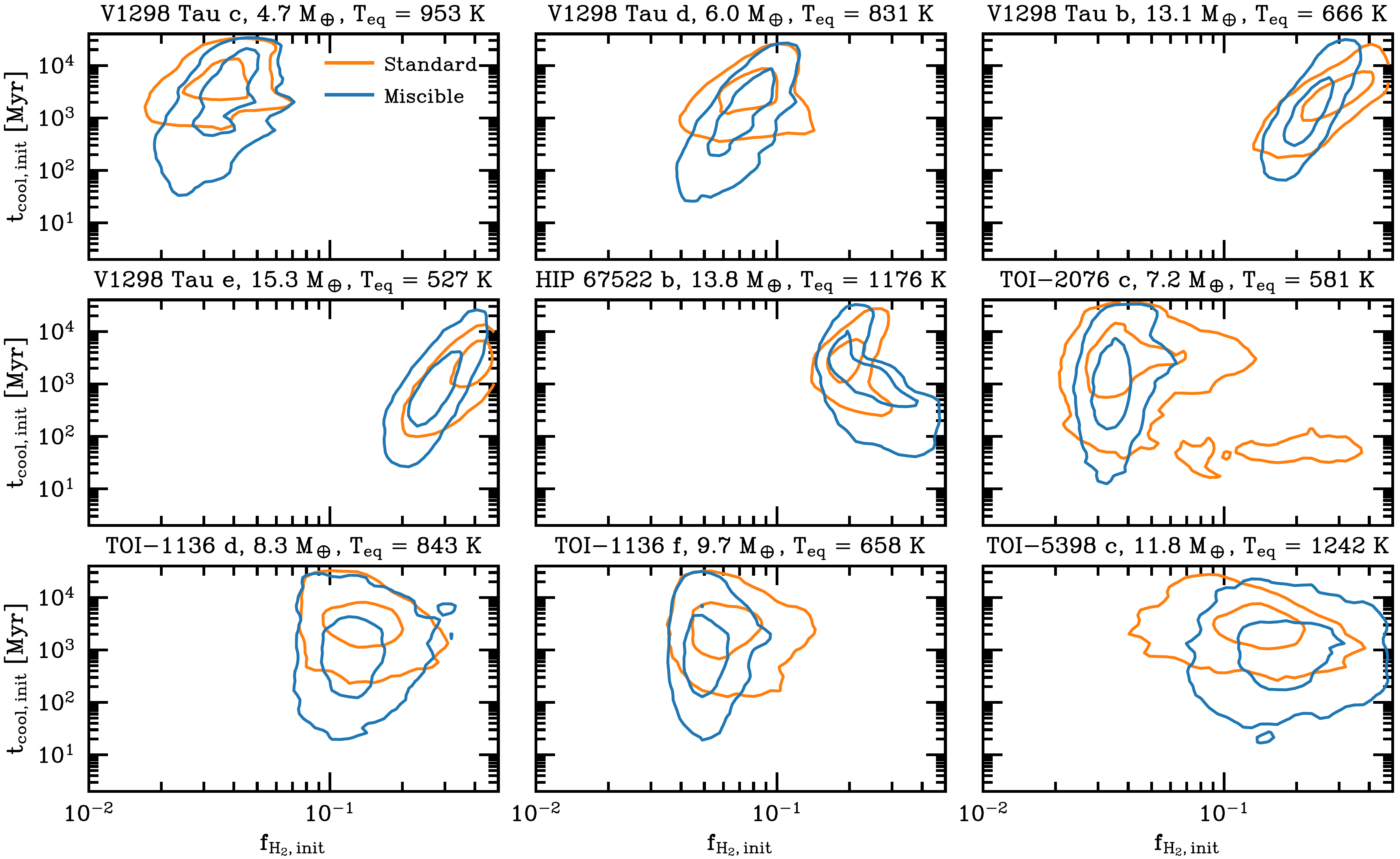}
    \centering
        \cprotect\caption{Marginalised posteriors for initial hydrogen mass fractions and initial cooling timescales for each planet in our sample shown in Table \ref{tab:planets}. Here we show $1\sigma$ and $2\sigma$ contours. Inferred distributions are shown for standard and miscible models in orange and blue respectively.} \label{fig:InitialConditions} 
\end{figure*}

\begin{figure*}
	\includegraphics[width=1.0\columnwidth]{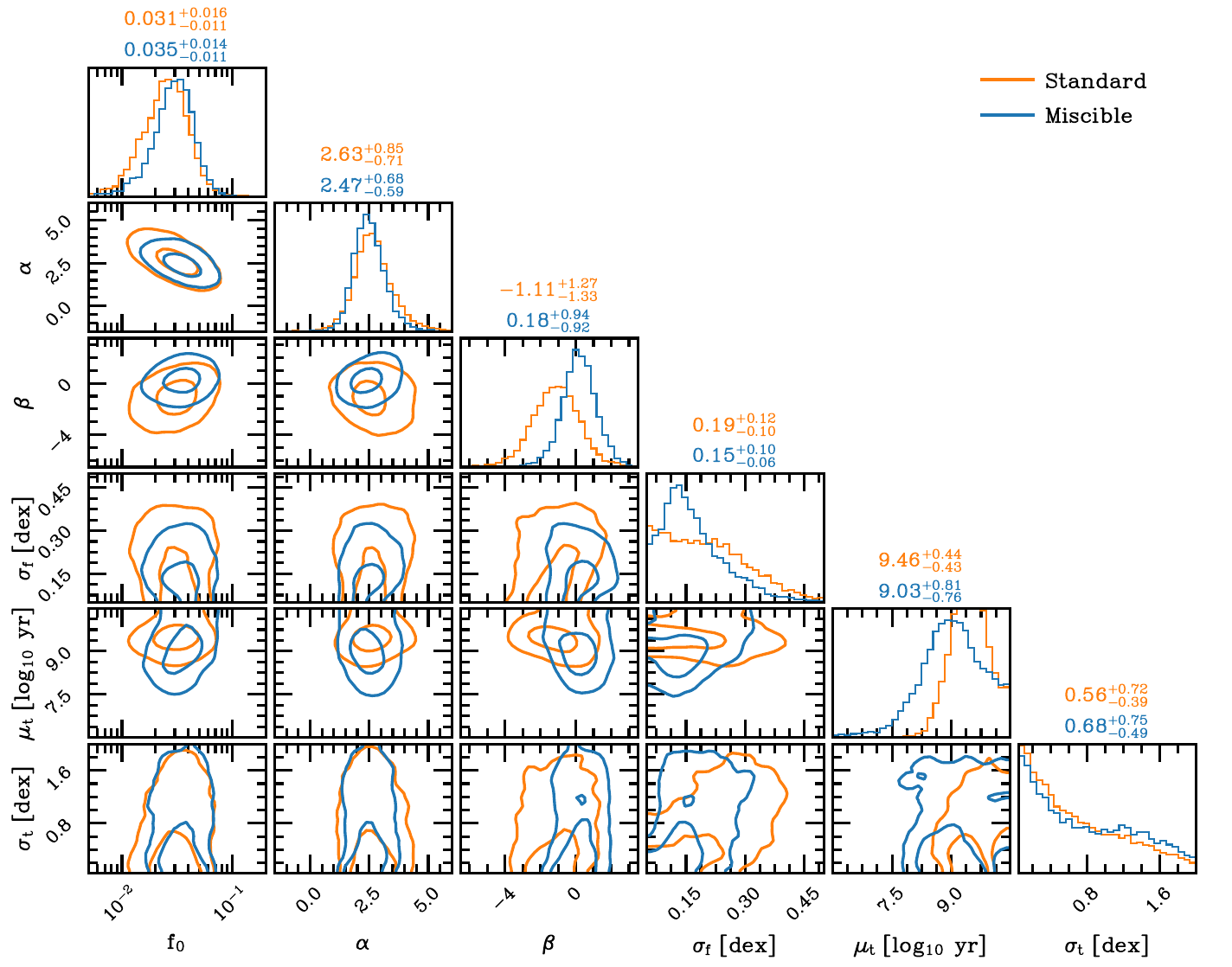}
    \centering
        \cprotect\caption{Marginalised posteriors for hyperparameters controlling the underlying distributions in initial hydrogen mass fractions and initial cooling timescales (see Equations \ref{eq:fH_init} and \ref{eq:tcool_init}). Here we show $1\sigma$ and $2\sigma$ contours. Inferred distributions are shown for standard and miscible models in orange and blue respectively.} \label{fig:HyperParameters} 
\end{figure*}

Here we present the posterior distributions for our Bayesian hierarchical inference model presented in Section \ref{sec:test_definitions} as applied to the $9$ observed planets in Table \ref{tab:planets}. Figure \ref{fig:InitialConditions} shows the marginalised posteriors for each planet's initial hydrogen mass fraction and initial cooling timescale. This is shown for standard and miscible models in orange and blue, respectively. In Figure \ref{fig:HyperParameters}, we show the marginalised posteriors for hyperparameters that control the population-level distributions for initial hydrogen mass fraction and initial cooling timescales in Equations \ref{eq:fH_init} and \ref{eq:tcool_init}.

\end{document}